\documentclass[%
 aip,
 amsmath,amssymb,
 reprint,%
]{revtex4-1}

\usepackage{graphicx}
\usepackage{dcolumn}
\usepackage{bm}
\usepackage{color}  

\usepackage[utf8]{inputenc}
\usepackage[T1]{fontenc}
\usepackage{mathptmx}

\begin{document}

\preprint{v. 1.5  21.04.20}

\title[Biquadratic exchange in FePS$_3$]{Evidence for biquadratic exchange in the quasi-two-dimensional antiferromagnet FePS$_3$}

\author{A. R. Wildes}
\email{wildes@ill.fr}
\affiliation{Institut Laue-Langevin, 71 avenue des Martyrs CS 20156, 38042 Grenoble Cedex 9, France}

\author{M. E. Zhitomirsky}
\affiliation{Univ. Grenoble Alpes, CEA, IRIG, PHELIQS, 17 avenue des Martyrs, 38000 Grenoble, France}

\author{T. Ziman}
\affiliation{Institut Laue-Langevin, 71 avenue des Martyrs CS 20156, 38042 Grenoble Cedex 9, France}
\affiliation{Univ. Grenoble Alpes, CNRS, LPMMC, 38000 Grenoble, France}

\author{D. Lan{\c{c}}on}
\altaffiliation{Present address: Paul Scherrer Institute, WHGA/150, 5232 Villigen PSI, Switzerland }
\affiliation{Institut Laue-Langevin, 71 avenue des Martyrs CS 20156, 38042 Grenoble Cedex 9, France}
\affiliation{Ecole Polytechnique F\'ed\'erale de Lausanne, SB ICMP LQM, CH-1015 Lausanne, Switzerland}

\author{H. C. Walker}
\affiliation{ISIS facility, Rutherford Appleton Laboratory, Harwell Oxford, Didcot OX11 0QX, UK}

\date{\today}

\begin{abstract}
FePS$_3$ is a van der Waals compound with a honeycomb lattice that is a good example of a two-dimensional antiferromagnet with Ising-like anisotropy.  Neutron spectroscopy data from FePS$_3$ were previously analysed using a straight-forward Heisenberg Hamiltonian with a single-ion anisotropy.  The analysis captured most of the elements of the data, however some significant discrepancies remained.  The discrepancies were most obvious at the Brillouin zone boundaries.  The data are subsequently reanalysed allowing for unequal exchange between nominally equivalent nearest-neighbours, which resolves the discrepancies.  The source of the unequal exchange is attributed to a biquadratic exchange term in the Hamiltonian which most probably arises from a strong magnetolattice coupling.  The new parameters show that there are features consistent with Dirac magnon nodal lines along certain Brillouin zone boundaries.
\end{abstract}

\maketitle


\section{\label{sec:Intro}Introduction}

FePS$_3$ belongs to a family of layered van der Waals compounds.  It is monoclinic with space group $C\frac{2}{m}$ and lattice parameters $a = 5.947$ {\AA}, $b = 10.3$ {\AA}, $c = 6.7222$ {\AA} and $\beta = 107.16^{\circ}$ \cite{Ouvrard85}.  The Fe$^{2+}$ ions are located at fractional coordinates $\left(0,0.3326,0\right)$, forming a slightly distorted honeycomb lattice in the $ab$ planes.  The compound is antiferromagnetic, ordering below its N{\'e}el temperature of $\sim$120 K \cite{Jernberg, Joy92} with a propagation vector of ${\bf{k}}_M = \left(01\frac{1}{2}\right)$ \cite{Kurosawa83, Lancon16}.  The high-spin ($S = 2$) Fe$^{2+}$ are aligned parallel to the $\bf{c^{\ast}}$ direction, i.e. normal to the $ab$ planes, forming ferromagnetic zig-zag chains that are parallel to the $a$ axis.  The chains are antiferromagnetically coupled along the $b$ and $c$ axes.  Figure \ref{fig:Structure}(a) shows a schematic of the magnetic structure in the $ab$ planes.  The paramagnetic susceptibility is highly anisotropic \cite{Joy92} and the compound appears to be a good example of an Ising-like antiferromagnet.

Other transition metals with 2+ valence may be substituted for iron, and the transition metal-PS$_3$ family of compounds has attracted considerable attention over the years \cite{Brec, Grasso, Susner, Wang}.  The weak van der Waals forces binding the $ab$ planes allow for intercalation with a wide variety of materials including lithium \cite{Brec, Grasso} and the compounds have been heavily investigated as prospects for battery materials.  The compounds are insulators, but they can be driven to become metallic on the application of pressure \cite{Coak19} providing avenues for fundamental studies of correlated electron physics.  One related compound, FePSe$_3$, can even be driven into a superconducting state at sufficiently high pressure \cite{WangY}.  The weak binding also extends to the magnetic coupling between the $ab$ planes, and the family are model systems for the study of low-dimensional magnetism.  FePS$_3$ is particularly interesting in this regard due to its Ising-like nature.  As shown by Onsager  \cite{Onsager}, long-ranged magnetic order is allowed in the two-dimensional Ising model, helping to explain the relatively high N{\'e}el temperature in FePS$_3$.  The discovery of graphene has led to a search for other compounds that may be delaminated to monolayer thickness.  The transition metal-PS$_3$ compounds are particularly interesting as they are magnetic \cite{Susner, Wang}.  They can be delaminated, and FePS$_3$ has been shown to maintain its Ising-like magnetic properties and antiferromagnetic ordering even for an isolated $ab$ plane \cite{Lee}, in agreement with the Onsager solution.    

Neutron scattering has been used to measure the spin wave dispersion and the data were fitted using linear spin-wave theory by Lan{\c{c}}on \emph{et al.} \cite{Lancon16}.   The spin-wave theory was based on a simple Heisenberg Hamiltonian with a single-ion anisotropy, and the honeycomb lattice was assumed to be perfect.  The data showed that FePS$_3$ was a good example of a two-dimensional antiferromagnet with very weak spin wave dispersion between the $ab$ planes, and a large energy gap of $\sim 15$ meV explaining its Ising-like behaviour.  Satisfactory fits to the data required the inclusion of up to the third nearest-neighbor in the $ab$ planes.  The fitting showed that  the interplanar exchange was $\sim 1/200$ times the magnitude of the nearest-neighbour intraplanar exchange and a large anisotropy dominating the magnitude of the gap.  However, there were some discrepancies between the calculations for the magnon dynamics and the measured neutron data.  The discrepancies were most apparent along the Brillouin zone boundaries where the data appeared to be far more symmetric about the central energy in the dynamic band width than could be reproduced by the model.

The data from Lan{\c{c}on} \emph{et al.} have consequently been reanalysed using a more general expression for the dynamic structure that allows for exchanges on nominally equivalent nearest-neighbours to be different.  The agreement with the data is significantly improved when fitting with this expression, although the results show large differences between the exchanges for nominally equivalent neighbours.  The differences appear to be too large to be explained by the slight departure from having a perfect honeycomb structure.  The differences are discussed in the context of adding a biquadratic exchange term, $\left({\bf{S}}_i\cdot{\bf{S}}_j\right)^2$, to the spin Hamiltonian.

\section{\label{sec:Intro}Hamiltonians and the dynamic structure factor}

\begin{figure*}
  \includegraphics[width=5in]{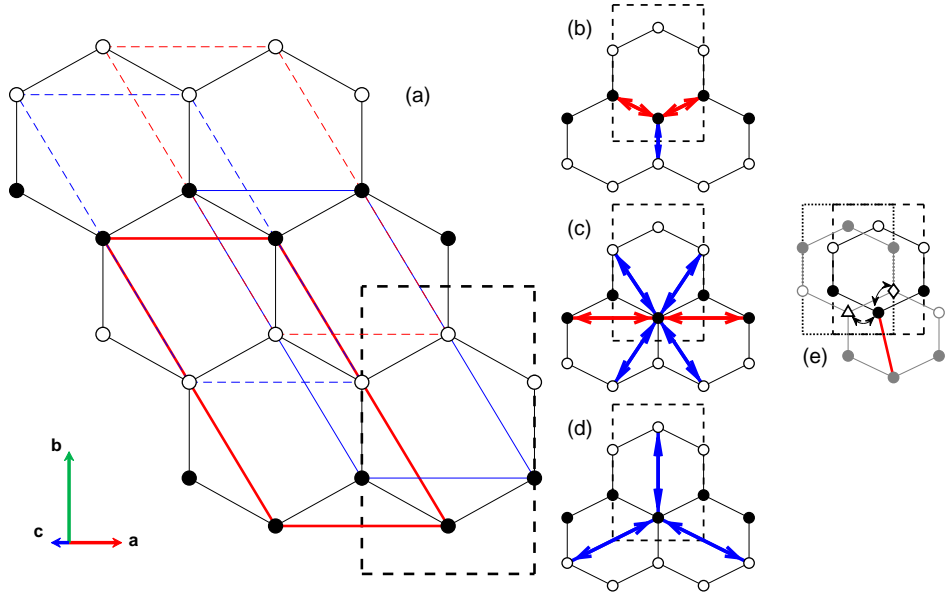}
  \caption{The magnetic structure and exchange parameters of FePS$_3$, as viewed along the ${\bf{c^\ast}}$ axis.  Open and closed symbols show moments into and out of the page respectively.  The monoclinic unit cell in the $ab$ planes is shown as black dashed lines.  (a). The magnetic structure in the $ab$ planes.  The magnetic structure may be decomposed into four interlocking Bravais lattices, shown as red and blue solid and dashed lines.  (b) the nearest-neighbour exchanges, $J_{1,\alpha}$ and $J_{1,\phi}$, indicated by blue and red arrows respectively.  (c) the second nearest-neighbour exchanges, $J_{2,\alpha}$ and $J_{2,\phi}$, indicated by blue and red arrows respectively.  (d) the third nearest-neighbour exchanges, $J_3$.  (e) the interlayer exchanges.  The moments shown as black circles are in an $ab$ plane one unit cell below the symbols representing all all other moments.  The exchange $J^{\prime}$ is indicated by the arrow between the open triangle and the closed circle, and $J^{\prime\prime}$ is indicated by the arrow between the open diamond and the closed circle.  The ${\bf{c}}_{mag}$ axis for the corresponding interlocking Bravais lattice shown in  (a) is shown as a  red solid line.}
  \label{fig:Structure}	
\end{figure*}

The Hamiltonian initially used to model the spin waves for FePS$_3$ was given by the equation:
\begin{equation}
  \hat{\mathcal{H}}=-\sum_{\left< ij \right>}J_{ij}{\bf{S}}_i\cdot{\bf{S}}_j-\Delta\sum_i\left(S^z_i\right)^2 .
  \label{eq:Hamiltonian_1}
\end{equation}
The dynamic structure factor was calculated using linear spin wave theory.  The magnetic structure was initial decomposed into four interlocking Bravais sublattices, shown in Figure \ref{fig:Structure}(a).  The axes for the magnetic Bravais lattices differ from those for the monoclinic unit cell, and the Miller indices for the two lattices are related via the transformation:
\begin{equation}
	\left[ \begin{array}{c} h \\ k \\ l\end{array}\right]
	= \left[\begin{array}{rcc} 1 & 0 & 0 \\ 1 & 1 & 0 \\ 0 & \frac{1}{2} & 1\end{array}\right]
	\left[ \begin{array}{c} h_{mag} \\ k_{mag} \\ l_{mag} \end{array}\right],
	\label{eqn:Matrix}
\end{equation}
where the subscript $mag$ refers magnetic Miller indices and the absence of a subscript refers to the monoclinic Miller indices.  The Holstein-Primakoff transformation was subsequently applied and, after Fourier transforming, the dynamic part of the Hamiltonian, ${\bf{H}_M}$, could be expressed in matrix form \cite{Wildes12, Lancon16}: 
\begin{equation}
  \label{eq:MatHamil}
  \begin{array}{ll}
  {\bf{H}_M}& = 2S\left[
    \begin{array}{llll}
      A & B^* & C & D^* \\
      B & A     & D & C \\
      C & D^* & A  & B^* \\
      D & C    & B  & A
    \end{array}
  \right].
  \end{array}
\end{equation}

The expressions for the matrix elements are given by the translation vectors between the different interlocking Bravais lattices.  The honeycomb lattices were assumed to be perfect, which is exactly true for the $C\frac{2}{m}$ unit cell in the limits that $b = \sqrt{3}a$ and the fractional position for the magnetic ions being at $\left(0,\frac{1}{3},0\right)$.  These conditions are very close to being met for FePS$_3$ \cite{Ouvrard85}.  The matrix elements in equation \ref{eq:MatHamil} then become expressions with rational fractions:
\begin{equation}
\label{eq:MatEl}
\begin{array}{ll}
A &=  2J_{2,\phi}\cos\left( 2\pi h_{mag}\right) -\Delta\\ 
   &~~~~~~ -2J_{1,\phi}+J_{1,\alpha} -  2J_{2,\phi} + 4J_{2,\alpha} +3J_3 + 2J^{\prime} + 2J^{\prime\prime} \\
    
B &= J_{1,\phi}\exp\left(\frac{2\pi i}{3}\left[2h_{mag}+\frac{k_{mag}}{2}\right]\right) \\
   &~~ \times\left(1+\exp\left(-2\pi ih_{mag}\right)\right) \\

C &= 2J_{2,\alpha}\left(\cos\left(\pi k_{mag}\right)+\cos\left(2\pi\left[h_{mag}+\frac{k_{mag}}{2}\right]\right)\right) \\
   &~~~~~~+J^{\prime}\cos\left(\pi\left[k_{mag}+2l_{mag}\right]\right)\\
   
D &=\exp\left(\frac{2\pi i}{3}\left[2h_{mag}+\frac{k_{mag}}{2}\right]\right) \\
   &~~\times\left\{
    \begin{array}{l}
    J_{1,\alpha}\exp\left(-2\pi i\left[h_{mag}+\frac{k_{mag}}{2}\right]\right) \\
   +J_3\left(
   \begin{array}{l}
   2\cos\left(\pi k_{mag}\right) \\
   + \exp\left(-2\pi i\left[ 2h_{mag}+\frac{k_{mag}}{2}\right]\right)
   \end{array}
   \right)
   \end{array}
   \right\} \\
  &~~+2J^{\prime\prime}\exp\left(\frac{\pi i}{3}\left[h_{mag}+k_{mag}\right]\right) \\
  &~~~~~~ \times\cos\left(\pi\left[h_{mag}+k_{mag}+2l_{mag}\right]\right)
\end{array}
\end{equation}

Figure \ref{fig:Structure}(b) shows the exchange interactions between the nearest-neighbours, of which there are three for a perfect honeycomb.  None of these neighbours are on the same Bravais lattice as the initial magnetic ion, with two of the neighbours on the ferromagnetically-coupled Bravais lattice and the third on one of the two antiferromagnetically-coupled Bravais lattices.  The exchanges between these neighbours are defined as $J_{1,\phi}$ and $J_{1,\alpha}$ for the ferro- and antiferromagnetically coupled moments respectively, shown as red and blue arrows in Figure \ref{fig:Structure}(b) respectively.  Inspection of Figure \ref{fig:Structure}(c) shows that a similar situation exists for the second nearest-neighbours, of which there are six on a perfect honeycomb.  Two of the neighbours are on the same Bravais lattice as the initial ion.  The interactions are shown as red arrows in the figure and they are defined as $J_{2,\phi}$.  The other four neighbours are on the other antiferromagnetically-coupled Bravais lattice, the exchanges are shown as the blue arrows and they are defined as $J_{2,\alpha}$.  Figure \ref{fig:Structure}(d) shows the third nearest-neighbour exchange, $J_3$, which are all between ions on the same antiferromagnetically-coupled Bravais lattice.

Figure \ref{fig:Structure}(e) shows the interplanar exchanges, along with the ${\bf{c}}_{mag}$ axis for one of the magnetic Bravais lattices.  There are two nearest neighbours on an adjoining layer, one on each of the antiferromagnetically-coupled Bravais lattices, and the two exchanges may be labelled $J^{\prime}$ and $J^{\prime\prime}$.  The interlayer distances to each of these nearest neighbours is identical in the limits of a perfect honeycomb lattice in the $ab$ planes and $a = -3c\cos\beta$ for the $C\frac{2}{m}$ unit cell.  

The expression for ${\bf{H}_M}$ given in equation \ref{eq:MatHamil} must be multiplied by a matrix to respect the commutation relations for the raising and lowering operators used in the Holstein-Primakov transformation.  The eigenvalues for the resulting matrix give the dispersion relations for the magnon energies, $\omega_{\bf{q}}$, through the expression:
\begin{equation}
\label{eq:Evals}
	\begin{array}{ll}
	\omega^2_{\bf{q}} = & A^2 + \left|B\right|^2-C^2-\left|D\right|^2 \\
	                       & \pm \left(4\left|AB^{\ast} - CD^{\ast}\right|^2 - \left|BD^{\ast} - DB^{\ast}\right|^2\right)^{\frac{1}{2}}.
	 \end{array}
\end{equation}
and the dynamic structure factor, $S_{mag}\left({\bf{Q}},\omega\right)$, is given by the normalized eigenvalues \cite{Wildes12}, which have been explicitly described for equation \ref{eq:MatHamil} by 
Wheeler \emph{et al.} \cite{Wheeler}.

For a perfect two-dimensional honeycomb lattice, the nearest-neighbour distances in figure \ref{fig:Structure}(b) are equivalent and there is no reason for $J_{1,\phi}$ and $J_{1,\alpha}$ to be inequivalent.  Equations \ref{eq:MatEl} may be simplified by equating $J_{1,\phi} = J_{1,\alpha} = J_1$, and similarly  $J_{2,\phi} = J_{2,\alpha} = J_2$.  The zig-zag magnetic structure shown in figure \ref{fig:Structure} is stablized due to the relative strengths of the three intraplanar exchange interactions, $J_1$, $J_2$ and $J_3$ \cite{Rastelli1, Rastelli2, Wildes20}.

The introduction of a biquadratic term into equation \ref{eq:Hamiltonian_1} causes the exchanges between different Bravais lattices to become inequivalent, i.e. $J_{1,\phi} \neq J_{1,\alpha}$ and $J_{2,\phi} \neq J_{2,\alpha}$.  The  Hamiltonian with biquadratic exchange takes the form \cite{Wysocki1, Wysocki2}:
\begin{equation}
  \hat{\mathcal{H}}=-\sum_{\left< ij \right>}\left[J_{ij}{\bf{S}}_i\cdot{\bf{S}}_j - \tilde{K}_{ij}\left({\bf{S}}_i\cdot{\bf{S}}_j\right)^2 \right] -\Delta\sum_i\left(S^z_i\right)^2.
  \label{eq:Hamiltonian_2}
\end{equation}
In the limit of a collinear antiferromagnetic structure, the biquadratic coupling between neighbours, $ \tilde{K}_{ij}$, may be incorporated into an effective exchange:
\begin{equation}
  \begin{array}{l}
    J_{ij,{\phi}} \rightarrow J_{ij} - 2\tilde{K}_{ij} S^2 \\
    J_{ij,{\alpha}} \rightarrow J_{ij} + 2\tilde{K}_{ij} S^2,  
 \end{array}
  \label{eq:HamilMod}
\end{equation}
where $J_{ij,{\phi}}$ describes the exchange between ferromagnetically-coupled neighbours, and $J_{ij,{\alpha}}$ is the exchange for antiferromagnetic neighbours.  The substitution in equation \ref{eq:HamilMod} is the result of the second-order expansion of the biquadratic term around the antiferromagnetic structure in terms of the bosonic magnon operators. Consequently, it is fully applicable to harmonic spin-wave theory and can be used to determine correctly the magnon dispersion and the corresponding dynamic structure factor. Equation \ref{eq:MatHamil} preserves its matrix form with precisely the same matrix elements as given by Equation \ref{eq:MatEl}.
However, the substitution is not appropriate for classical energy calculations which must be performed using the full spin Hamiltonian in equation \ref{eq:Hamiltonian_2}.


 \section{Fitting the neutron spectroscopy data}
 
\begin{figure*}
  \includegraphics[width=5in]{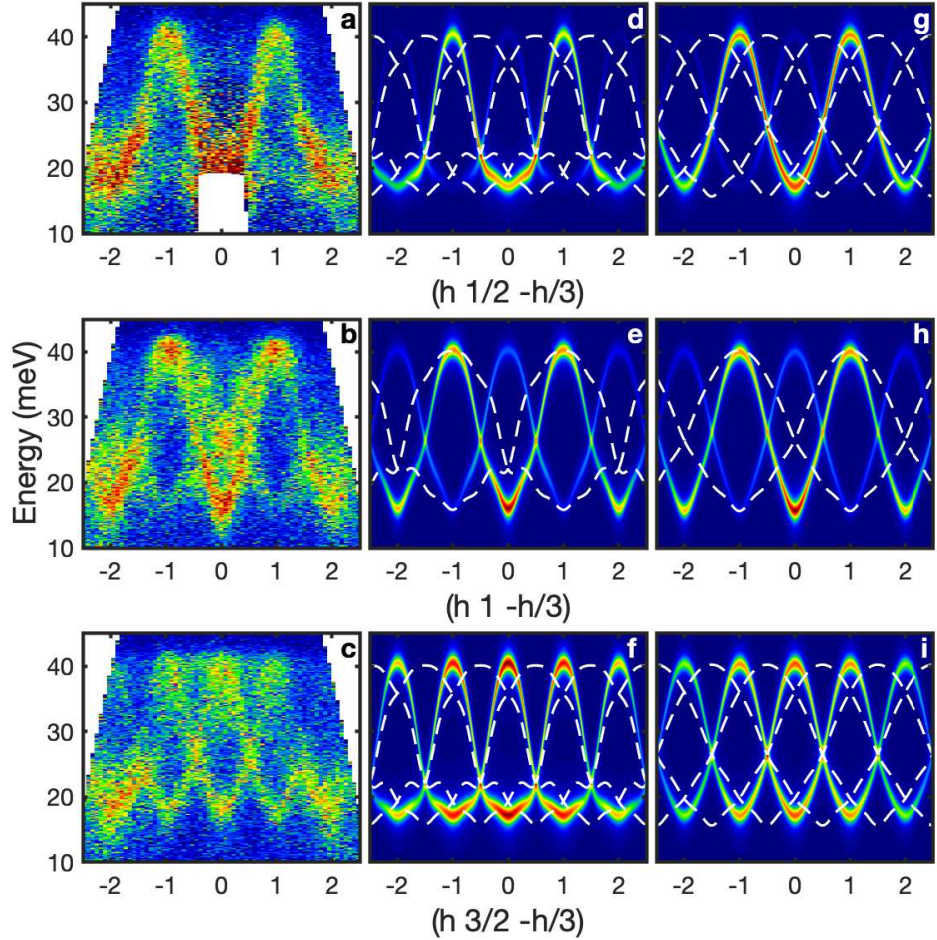}
  \caption{Neutron scattering data and fit results for $S\left({\bf{Q}},\omega\right)$ along trajectories parallel to $\left(h0\frac{\overline{h}}{3}\right)$ for FePS$_3$.  Parts (a)-(c) show the neutron scattering data measured at 5 K from the experiments performed by Lan{\c{c}}on \emph{et al.} \cite{Lancon16}.  Parts (d)-(f) show the calculated $S\left({\bf{Q}},\omega\right)$ using the fitted parameters on constraining $J_{1,\phi} = J_{1,\alpha}$ and $J_{2,\phi} = J_{2,\alpha}$, following the analysis by Lan{\c{c}}on \emph{et al.} \cite{Lancon16}. Parts (g)-(i) show the calculations from fits where the corresponding nearest-neighbour exchanges were not constrained to be equal.  Trajectories centred at $\left(0\frac{1}{2}0\right)$ are shown in parts (a), (d) and (g), trajectories centred at $\left(010\right)$ are show in parts (b), (e) and (h) and trajectories centred at $\left(0\frac{3}{2}0\right)$ are shown in parts (c), (f) and (i).  The dashed lines in parts (d)-(i) show the calculated positions for the contributions from crystal twins.  Figures (a)-(f) are adapted with permission from Physical Review B, 94, 214407 (2016) (reference [\cite{Lancon16}]), copyrighted 2016 by the American Physical Society.}
  \label{fig:h0h3}	
\end{figure*}

\begin{figure*}
  \includegraphics[width=5in]{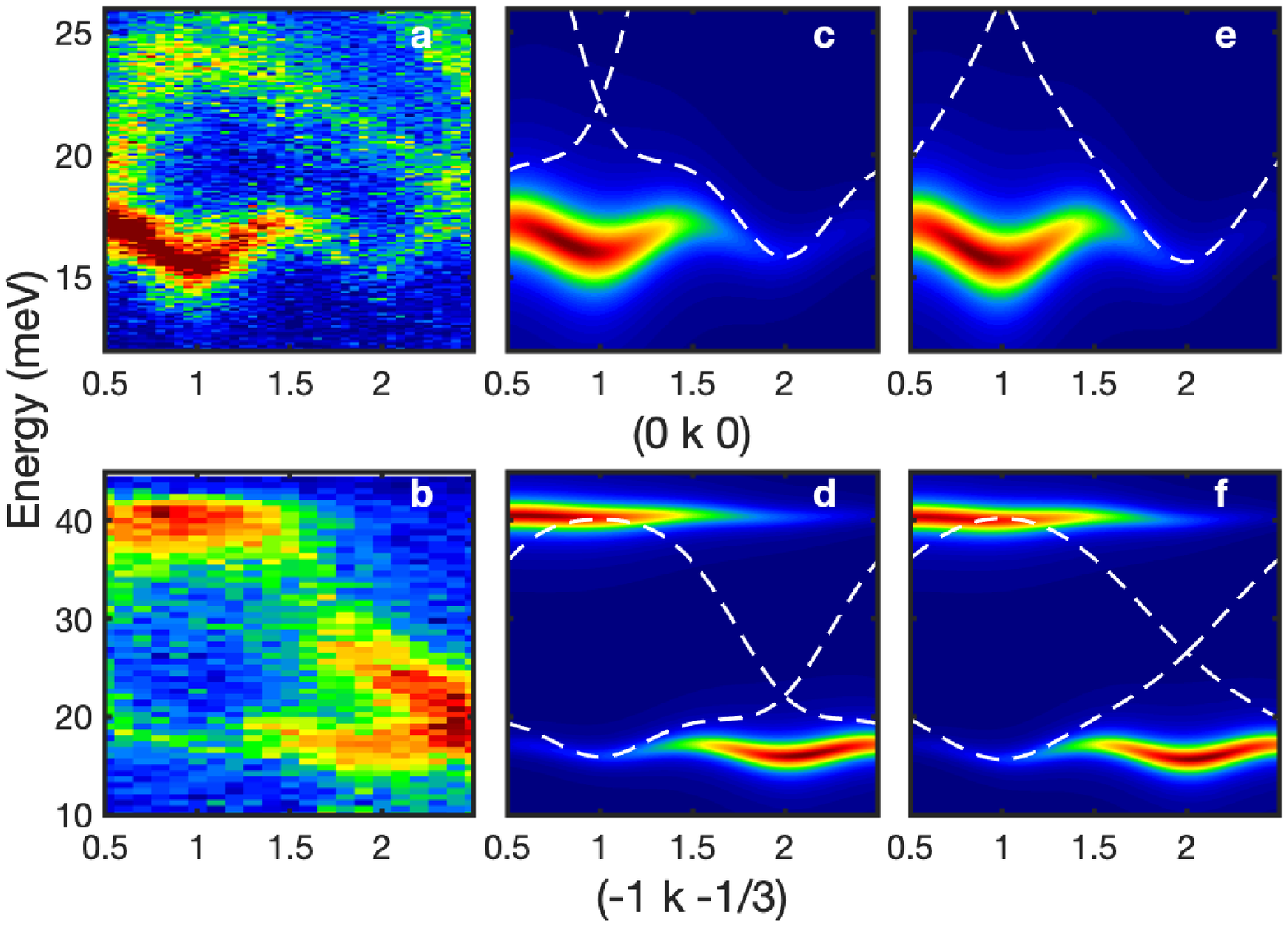}
  \caption{Neutron scattering data and fit results for $S\left({\bf{Q}},\omega\right)$ along trajectories parallel to $\left(0k0\right)$ for FePS$_3$.  Parts (a)-(b) show the neutron scattering data measured at 5 K from the experiments performed by Lan{\c{c}}on \emph{et al.} \cite{Lancon16}. Parts (c)-(d) show the calculated $S\left({\bf{Q}},\omega\right)$ using the fitted parameters on constraining $J_{1,\phi} = J_{1,\alpha}$ and $J_{2,\phi} = J_{2,\alpha}$, following the analysis by Lan{\c{c}}on \emph{et al.} \cite{Lancon16}.  Parts (e)-(f) show the calculations from fits where the corresponding nearest-neighbour exchanges were not constrained to be equal.  Trajectories along $\left(0k0\right)$ are shown in parts (a), (c) and (e) and trajectories along $\left(\overline{1}k\frac{\overline{1}}{3}\right)$ are show in parts (b), (d) and (f).  The dashed lines in parts (d)-(f) show the calculated positions for the contributions from crystal twins. Figures (a) and (c) are adapted with permission from Physical Review B, 94, 214407 (2016) (reference [\cite{Lancon16}]), copyrighted 2016 by the American Physical Society.}
  \label{fig:0k0}	
\end{figure*}

\begin{figure*}
  \includegraphics[width=5in]{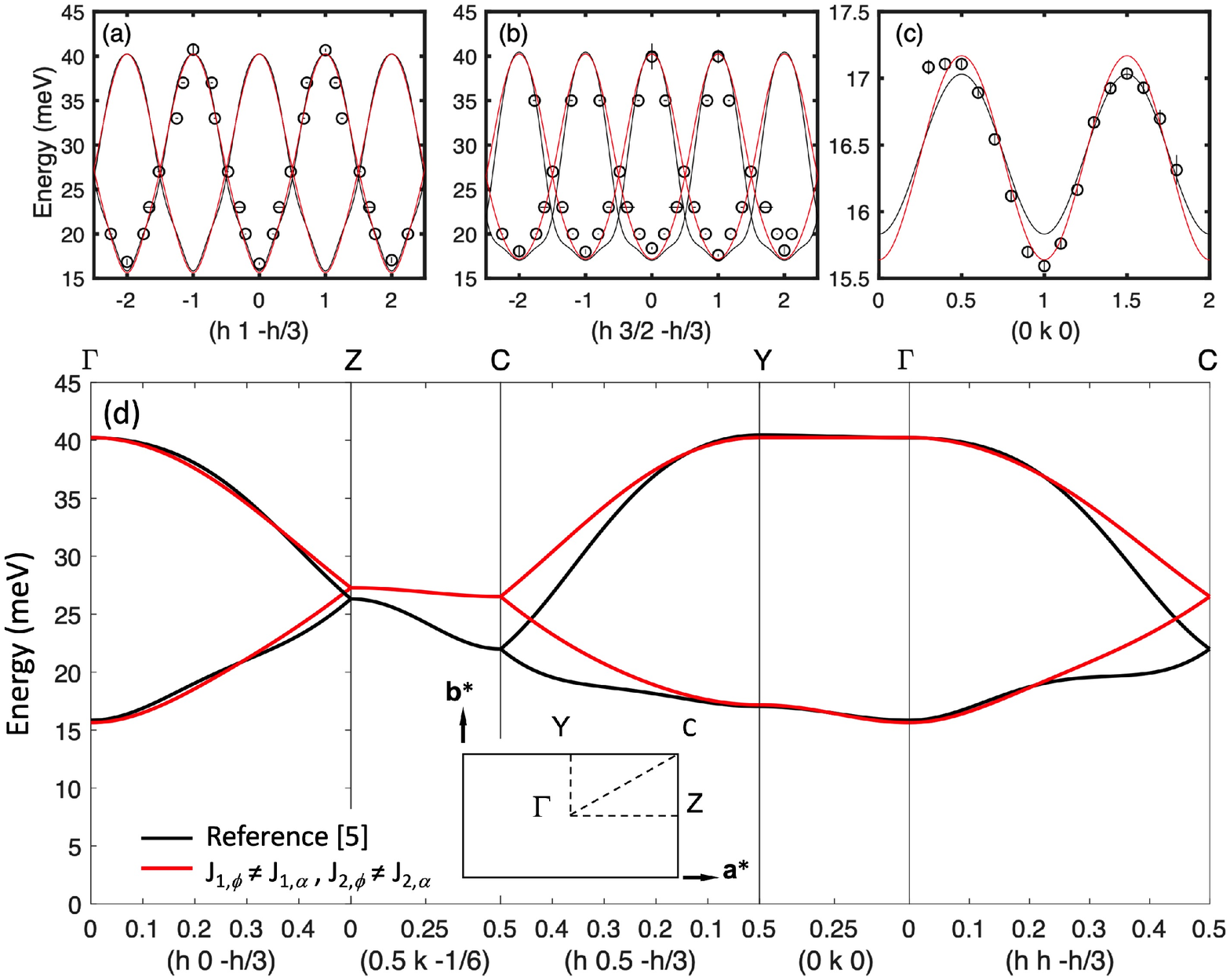}
  \caption{The fitted data and dispersion curves for slices along different reciprocal space directions for FePS$_3$.  The black lines show the dispersions following the analysis from Lan{\c{c}}on \emph{et al.} \cite{Lancon16}.  The red lines show the dispersions after allowing for a biquadratic interaction in the fitting, i.e. $J_{1,\phi} \neq J_{1,\alpha}$ and $J_{2,\phi} \neq J_{2,\alpha}$.  The data points in (a)-(c) were determined by taking linear cuts through figures \ref{fig:h0h3}(b) and (c) and figure \ref{fig:0k0}(a) respectively.  (d) The calculated dispersion over a selection of reciprocal space directions, with the insert showing the Brillouin zone in the $\left(a^*b^*\right)$ plane.}
  \label{fig:Disp}	
\end{figure*}

Lan{\c{c}}on {et al.} \cite{Lancon16} used the MERLIN time-of-flight neutron spectrometer at the ISIS facility, UK, to measure the magnetic scattering from FePS$_3$ at 5 K.  The instrument captures a large volume of reciprocal space, and the data may subsequently be presented as slices.  The data are shown in figures \ref{fig:h0h3}(a)-(c) and \ref{fig:0k0}(a)-(b), corresponding to slices along different directions in the \emph{ab} planes.  The insert of figure \ref{fig:Disp}(d) shows a schematic for the Brillouin zone for a two-dimensional zig-zag structure on a honeycomb lattice.  The ${\bf{a}}^{\ast}$ and ${\bf{b}}^{\ast}$ vectors marked on the Brillouin zone are for the $C\frac{2}{m}$ reciprocal lattice.

Figures \ref{fig:h0h3} show slices parallel to the monoclinic $\left(h0\frac{\overline{h}}{3}\right)$ direction.  The slices are centred at different values of $k$, with figure \ref{fig:h0h3}(b) following the $\Gamma-$Z axis through Brillouin zone centres and figures \ref{fig:h0h3}(a) and (c) being slices along Y$-$C zone boundaries.  Figures \ref{fig:0k0} show slices parallel to the ${\bf{b}}^{\ast}$ axis, with figure \ref{fig:0k0}(a) showing the data along the $\Gamma-$Y axis through Brillouin zone centres and figure \ref{fig:0k0}(b) showing the slices along Z$-$C Brilliouin zone boundaries.  The data clearly show two intersecting spin wave branches, as anticipated from equation \ref{eq:Evals}.  The branches are highly dispersive along the $\left(h0\frac{\overline{h}}{3}\right)$ direction and only weakly dispersive along directions parallel to the ${\bf{b}}^{\ast}$ axis.  The data are complicated by the presence of twin domains prevalent in this family of compounds \cite{Murayama}.  The contributions from all the domains are superimposed in the data, however they can be differentiated as they are quite distinct and are related by well-defined rotation matrices \cite{Lancon16}.  The data also showed a small amount of spurious scattering, seen primarily in figure \ref{fig:0k0}(a) as an arc of intensity descending from $\sim 25 $ meV at $k=0.5$ to $\sim 18$ meV at $k = 2.5$ and as a small amount of extra intensity in the top right-hand corner of the figure.

Spin wave energies were extracted from figures \ref{fig:h0h3}(b) and (c) and figure \ref{fig:0k0}(a) by Lan{\c{c}}on \emph{et al.}.  One-dimensional cuts were taken through the data and the energies were determined by fitting damped harmonic oscillator functions, taking care to avoid or account for the spurious scattering and the contributions from other magnetic domains.  The extracted spin wave energies are shown in figures \ref{fig:Disp}(a)-(c).  The energies were then simultaneously fitted with equation \ref{eq:Evals} assuming that $J_1 = J_{1,\phi} = J_{1,\alpha}$, $J_2 = J_{2,\phi} = J_{2,\alpha}$ and $J^{\prime} = J^{{\prime}{\prime}}$, and the fits are shown as black lines in figures \ref{fig:Disp}(a)-(c).  The exchange parameters for the three nearest intralayer neighbours are shown in table \ref{tab:Exchange}.  The nearest-neighbour exchange, $J_1$, is ferromagnetic, the second-nearest-neighbour exchange, $J_2$, is very close to zero, and the zig-zag magnetic structure for FePS$_3$ is stabilized by a relatively strong antiferromagnetic $J_3$ \cite{Rastelli1, Wildes20}.  The interlayer exchange, determined from separate data measured using a three-axis spectrometer \cite{Lancon16}, was found to be very small, $J^{\prime} = -0.0073\left(3\right)$ meV.  Its inclusion has only a very minor effect on modelling the spin waves and it will be ignored in subsequent discussion.
 
\begin{table*}
  \caption{\label{tab:Exchange}Exchange parameters and anisotropy determined from fitting equation \ref{eq:Evals} with various constraints on the matrix elements in equations \ref{eq:MatEl}.  All the fit parameters are expressed in meV.  The fit quality, given by $\chi^2$, is also tabulated for the respective fits,}
  \begin{tabular}{l|c|c|c|c|c|c|c|}
  																				& $\Delta$ & $J_1$	& $2\tilde{K}_1 S^2$ 	& $J_2$		& $2\tilde{K}_2 S^2$	& $J_3$	& $\chi^2$ \\
\hline
  $J_{1,\phi} = J_{1,\alpha}$, $J_{2,\phi} = J_{2,\alpha}$ (c.f. Lan{\c{c}on \emph{et al.}}\cite{Lancon16})		& 2.66	 & 1.46	&					    	&  $-0.04$		& 					     	& $-0.96$	& 147	 \\
  $J_{1,\phi} \neq J_{1,\alpha}$, $J_{2,\phi} = J_{2,\alpha}$									& 2.53	 & 0.9	& 0.55					& $-0.047$	& 						& $-0.64$	& 57.7	 \\
  $J_{1,\phi} = J_{1,\alpha}$, $J_{2,\phi} \neq J_{2,\alpha}$									& 2.46	 & 1.46	&						& $-0.22$		& 0.2						& $-0.55$ & 57.4	 \\
  $J_{1,\phi} \neq J_{1,\alpha}$, $J_{2,\phi} \neq J_{2,\alpha}$									& 2.52	 & 0.98	& 0.46					& $-0.155$	& 0.085					& $-0.45$	& 49		\\
  \end{tabular} 
\end{table*} 

The dynamic structure factors were calculated with these parameters and they are shown in figures \ref{fig:h0h3}(d)-(f), corresponding to the data in figures \ref{fig:h0h3}(a)-(c) respectively, and figures \ref{fig:0k0}(c)-(d), corresponding to the data in figures \ref{fig:0k0}(a)-(b) respectively.  The expected spin wave contributions from the twinned domains are also plotted in these figures as white dashed lines.  Inspection of the figures shows reasonable agreement between the neutron data and the calculated $S_{mag}\left({\bf{Q}},\omega\right)$, however there are numerous discrepancies.  The data and the model agree particularly well between figures \ref{fig:h0h3}(b) and (e), showing slices through Brillouin zone centres.  The same is true when comparing the data and models in figures \ref{fig:0k0}(a) and (c), and figures \ref{fig:0k0}(b) and (d).  However. comparison between the data and the model in figures \ref{fig:h0h3}(a) and (d) and figures \ref{fig:h0h3}(c) and (f) are not as good.  The data are far more symmetric about the mid-point of the dynamic range for the spin waves at $\sim 27$ meV than the model calculations.  The lowest energy spin waves, at $\sim 15$ meV at the Brillouin zone centres, are more rounded in the modelling than are apparent in the data.  Furthermore, while both the data and the model in figures \ref{fig:h0h3}(c) and (f) show two intersecting spin wave branches, the point of intersection in the data is at $\sim 27$ meV while it is at a lower energy of $\sim 20$ meV in the model. 

Discrepancies are also apparent on comparing the expected contributions from the twinned domains.  These contributions are most clearly seen in figure \ref{fig:h0h3}(b) as dispersive modes that cross at $h$=0  and $\omega = 27$ meV, in figure \ref{fig:0k0}(a) as a steeply rising dispersive mode originating at $k = 0.5$, and a mode in figure \ref{fig:0k0}(b) descending from $\sim 30$ meV at $k \sim 1.7$ to 17 meV at $k = 2.5$.  These contributions are not reproduced well by the modelling, which tends to shift them to lower energies.

The fits were substantially improved by allowing for $J_{1,\phi} \neq J_{1,\alpha}$ and/or $J_{2,\phi} \neq J_{2,\alpha}$.  Table \ref{tab:Exchange} shows the fit parameters, presented in a form consistent with equation \ref{eq:HamilMod}.  The table includes a column showing a goodness-of-fit expression, $\chi^2$, for the different fits to the spin waves.  Three fits with different non-equal exchanges were performed in total.  The value for $\chi^2$ decreased substantially by a factor $\sim 3$ on allowing either $J_{1,\phi} \neq J_{1,\alpha}$ or $J_{2,\phi} \neq J_{2,\alpha}$ while constraining the other exchanges to be equal.  Allowing both to be unequal further decreased $\chi^2$ by $\sim 15 \%$.  The last fits are shown as red lines in figures \ref{fig:Disp}(a)-(c) and the calculated $S\left({\bf{Q}},\omega\right)$ for the corresponding parameters are shown for the respective slices through reciprocal space in figures \ref{fig:h0h3}(g)-(i) and figures \ref{fig:0k0}(e)-(f), although it was barely possible to differentiate between similar plots using the parameters when either the first or second nearest-neighbour exchanges were constrained to be equal.  Inspection of the figures shows a much better agreement between the data and the calculations with the new parameters.  The same is true when comparing the contributions from the twinned domains in figures \ref{fig:h0h3} and \ref{fig:0k0}.

Figure \ref{fig:Disp}(d) shows the calculated spin wave dispersions over the whole Brillouin zone in the $\left({\bf{a}}^{\ast}, {\bf{b}}^{\ast}\right)$ planes for both the parameters determined by Lan{\c{c}}on \emph{et al.} and the parameters from the new fits.  The biggest differences are seen along the Z$-$C Brillouin zone boundary, where the spin waves become almost dispersionless when the exchange interactions for equivalent nearest-neighbour distances are not constrained to be equal. 

Comparison of the values in table \ref{tab:Exchange} shows that the magnitude of the single-ion anisotropy does not vary greatly between the fits.  This is expected, as the anisotropy is the dominant factor behind the observed spin wave gap of $\sim 15$ meV and is not multiplied by a {\bf{q}}-dependent term in equation \ref{eq:MatEl}.  Allowing for non-equal $J_1$ resulted in substantially different values, with one of the two exchange parameters being close to that determined by Lan{\c{c}}on \emph{et al.} and the other being significantly smaller by $\approx 1$ meV.  The same is true when fitting with non-equal $J_2$, with a smaller absolute difference between the two parameters of $0.2 - 0.4$ meV but a much larger relative difference.  The value for $J_3$ decreased by $\approx 30$ {\%} for non-equal $J_1$ and/or $J_2$.  

\section{Discussion}
The constraints implicit in the assumptions adopted by Lan{\c{c}}on \emph{et al.} in their analysis\cite{Lancon16} were that the lattice parameters were related through the expressions $b=\sqrt{3}a$ and $a=-3c\cos\beta$ and that the fractional coordinates for the Fe$^{2+}$ ions were $\left(0,\frac{1}{3},0\right)$, meaning that the magnetic lattice formed a perfect honeycomb and that the exchange pathways were identical for all corresponding nearest-neighbours.  Inspection of the crystal structure for FePS$_3$ shows this to be a very good assumption \cite{Ouvrard85}, with the constraints on the lattice parameters being correct to better than 0.05 {\%} and the constraint on the fractional coordinates being correct to better that 3 {\%}.  The assumptions also require that the exchange pathways, mediated by the sulfur atoms and with a possible contribution by the phosphorus, between nominally equivalent nearest-neighbours are also equal.  Again, inspection of the crystal structure shows this to be a good assumption.  FePS$_3$ is only very slightly distorted from being an antiferromagnet with a perfect honeycomb lattice.

Equations \ref{eq:MatEl} require that the Fe$^{2+}$ ions are on a perfect honeycomb in that the vectors connecting the interlocking Bravais lattices are constructed from rational fractions of the magnetic basis vectors.  Nonetheless, they will allow for deviations from a perfect structure in that the exchanges, shown in figure \ref{fig:Structure}, connect Fe atoms with equivalent nearest-neighbour distances in the distorted honeycomb.  A difference between, for example, $J_{1,\phi}$ and $J_{1,\alpha}$ in the fits could reflect differences in the nearest-neighbour distances and small distortions in the bond angles with the sulfur atoms that would mediate the exchange.  However, the degree of the structural distortion would seem to be far too small to result in a difference of $\approx 1$ meV observed in table \ref{tab:Exchange}.  The assumption is supported by ab-initio calculations \cite{Chittari,Gu19} which calculated magnetic exchange parameters for energetically-relaxed structures.  Consistent with observations, the calculated structures were only very slightly distorted and the calculations gave unique exchange parameters for nominally equivalent nearest-neighbours which, depending on the magnitude of the on-site repulsion term, were comparable to those of Lan{\c{c}}on \emph{et al.}

It appears more likely that there is another important energy term in the magnetic Hamiltonian for FePS$_3$.  The biquadratic term is a suitable candidate for allowing differences between nominally equivalent nearest-neighbours through the introduction of the biquadratic exchange parameter, $\tilde{K}$.  Biquadratic exchange interactions can occur in Mott insulators with spins $S \geq 1$ from fourth-order perturbation processes in electron hopping \cite{Mila, Tanaka}.  The biquadratic coupling may be further enhanced by metallicity \cite{Wysocki1, Wysocki2} or by substantial  magnetoelastic interactions \cite{Penc}.  The latter effect is particularly applicable for FePS$_3$.   The onset of magnetic order at the N{\'e}el temperature is accompanied by a discontinuous change in the temperature-dependence of the lattice parameters \cite{Murayama, Jernberg} leading to the conclusion that the transition is of first order.  Furthermore, FePS$_3$ shows strong magnetoelastic effects at high magnetic fields to the point where a sample will self-delaminate when a sufficiently high field is applied along the ${\bf{b}}$ direction \cite{Wildes20}.  Consequently, the magnetoelastic effect may be responsible for the large values of the biquadratic constant $\tilde{K}$. found from fitting the experimental magnon dispersion.

The inclusion of the biquadratic term does not lead to a complete understanding of the magnetic dynamics of FePS$_3$.  The high-field magnetization measurements showed a high degree of anisotropy when the field was applied in the $ab$ planes, being normal to the ordered moment direction \cite{Wildes20}, which would not be observed for dynamics governed by equation \ref{eq:Hamiltonian_2}.  The observed anisotropy is presumed to be due to the specifics of the coupling between the spin wave and phonon dynamics which are not captured with the simple biquadratic term.  

Furthermore, application of a sufficiently high magnetic field along the ordered moment direction led to an intermediate partially-magnetized ordered state before being driven to a fully magnetized state.  Inspection of a calculated phase diagram for an Ising two-dimensional honeycomb system with exchange interactions up to the third-nearest neighbour showed that the exchange parameters from Lan{\c{c}}on \emph{et al.} would not give rise to such a partially-magnetized state \cite{Wildes20}, which requires that $J_2 \leq -J_1/2$.  The inclusion of the biquadratic term will not solve this discrepancy.  Not only do the parameters in table \ref{tab:Exchange} show that $J_{2,\{\alpha,\phi\}} > -J_{1,\{\alpha,\phi\}}/2$ for all combinations of $\{\alpha,\phi\}$, the effective exchange given in equation \ref{eq:HamilMod} is only applicable for linear spin waves and must not be used to calculate the energies of the magnetic structures.  The proposed solution to the discrepancy was to adapt equation \ref{eq:Hamiltonian_1} to become an XXZ Hamiltonian \cite{Wildes20}.  The exchange parameters found from neutron spectroscopy would be consistent with the exchange parameters acting on the $S^x_iS^x_j$ product, while the magnetic structure would be determined with the exchange parameters acting on $S^z_iS^z_j$.  The solution is coherent with the introduction of a biquadratic term considered here as only the spin wave dynamics are considered, although future efforts will be made to investigate the need for an anisotropic $\tilde{K}$ in the Hamiltonian.

There is some ambiguity in the fits concerning whether unequal exchanges are required for $J_1$ only, for $J_2$  only, or for both.  The quality of the fits is essentially the same if either $J_1$ or $J_2$ are allowed to be unequal, and it is this allowance that gives the greatest improvement over the findings from Lan{\c{c}}on \emph{et al.}.  Allowing both to be unequal did improve the quality of the fits, but not so much as to be conclusive.  It is likely that $J_1$ is the most important exchange to be unequal.  The exchange paths between the Fe$^{2+}$ ions will be mostly mediated by the sulfur atoms.  Inspection of the crystal structure shows that the exchange paths between nearest-neighbours are straight forward, but that exchange paths to the second nearest-neighbours are convoluted  \cite{LeFlem,Lancon18}.  It seems probable that $J_2$ remains close to zero and that $J_1$, which is strong, is the exchange most affected.

A final observation is of interest in the spin dynamics for FePS$_3$. The data, and now the modelling, establish that the spin waves along the Z$-$C boundary consist of two approximately linear branches that cross.  As shown in figure \ref{fig:Disp}(d), the crossing point is almost dispersionless, varying by $\sim 1$ meV from the Z to the C points.  The spin wave dispersion is thus consistent with the presence of two-dimensional Dirac magnon nodal-lines, as described by Owerre \cite{Owerre}.  Dirac points, defined as a point at which valence and conduction bands for a two-dimensional conductor form cones that touch at their apexes, feature in the description of the electron transport in graphene.  The points have an associated chirality to their Berry curvature that leads to magnetoelectric effects that can be exploited in technology.  Recent studies have searched for magnetic analogues in the spin dynamics of layered honeycomb ferromagnets such as CrX$_3$ (X = Br, Cl, I) \cite{Persh}, which may lead to applications in spintronics.  The Dirac magnon points in these compounds are found in the Brillouin zone corners, where two-dimensional magnon branches with linear dispersions cross.  Owerre has proposed that these Dirac points can be generalised to nodal lines and have associated topological characteristics.  The lines give rise to a peak in the magnon density of states.  The density of states has been measured for FePS$_3$ \cite{Wildes12}.  Close inspection of the data shows a small feature at slightly less than 30 meV, corresponding to the energies of the magnon modes intersecting along the Z$-$C boundary, consistent with the prediction of Owerre.  More detailed studies of the magnetic dynamics in FePS$_3$ are thus of interest for possible topological effects.

\section{Conclusions}
Neutron spectroscopy data showing the spin waves in FePS$_3$ have been reanalysed, allowing for unequal exchange parameters between nominally equivalent nearest-neighbours.  The results show that substantial differences arise between ferromagnetically- and antiferromagnetically-coupled neighbours.  The differences appear to be too large to be due to structural distortions, as FePS$_3$ is very close to having a perfect honeycomb lattice, and instead can be explained with the introduction of a biquadratic exchange term into the Hamiltonian.   The biquadratic term causes nominally equivalent sites to become inequivalent.  It can be attributed to a strong magnetoelastic coupling, which causes the magnetic ordering transition to be of first order and is in evidence in magnetometry experiments, where the magnetostriction is so strong that the sample will delaminate when a sufficiently high field is applied along the {\bf{b}} axis.

\section{Data availability}
Raw data were generated by the MERLIN instrument at the ISIS Neutron and Muon Source, UK (experiment RB1410242).  Derived data supporting the findings of this study are available from the corresponding author upon reasonable request.

\begin{acknowledgments}
The neutron experiment at the ISIS Neutron and Muon Source was supported by a beam time allocation from the Science and Technology Facilities Council (UK).  D.L. acknowledges funding by The Swiss National Science Foundation Grants No. 146870 and No. 166298 and Sinergia network MPBH Grant No. 160765.  D. L. and A. R. W. thank Professor H. R{\o}nnow for stimulating discussions.
\end{acknowledgments}



\bibliography{MPS3}

\end{document}